\pdfoutput=1
\documentclass[a4paper,12pt]{article}
\usepackage[margin=2.5cm]{geometry}
\usepackage{authblk}

\usepackage{amsmath}
\usepackage{slashed}
\usepackage{indentfirst}
\usepackage[colorlinks]{hyperref}
\usepackage[subentry=false,style=numeric-comp,sortcites=false,sorting=none]{biblatex}
\usepackage{tikz}
\usetikzlibrary{intersections}
\usetikzlibrary{patterns}
\usetikzlibrary{arrows,decorations.pathmorphing,backgrounds,positioning,fit,petri}
\usetikzlibrary{decorations.markings}
\usetikzlibrary{mindmap}
\usetikzlibrary{calc}
\usetikzlibrary{arrows}
\usetikzlibrary{intersections,through,hobby}
\definecolor{bla}{HTML}{03396C}
\definecolor{blaa}{HTML}{005B96}
\definecolor{blaaa}{HTML}{6497B1}

\hypersetup{
  colorlinks, 
  bookmarksopen, 
  bookmarksnumbered,
  citecolor=blaa, 		
  linkcolor=blaa,    	
  urlcolor=blaa,			
}

\newcommand{\ep}{\ensuremath{\varepsilon}}
\newcommand{\I}{\ensuremath{{\mathrm{i}}}}
\newcommand{\al}{\ensuremath{\alpha}}
\newcommand{\D}{\ensuremath{\mathrm{d}}}
\newcommand{\nf}{\ensuremath{n_f}}
\newcommand{\eg}{{\it e.g.}}
\newcommand{\ie}{{\it i.e.}}

\title{\Large \bf \color{bla} Four-loop singularities \\ of the massless fermion propagator \\
  in quenched three-dimensional QED}

\author[1]{A.\ F.~Pikelner}
\author[2]{V.\ P.~Gusynin}
\author[1]{A.\ V.~Kotikov}
\author[3]{S.~Teber}

\affil[1]{Bogoliubov Laboratory of Theoretical Physics, Joint Institute for Nuclear Research, 141980 Dubna, Russia.}
\affil[2]{Bogolyubov  Institute  for  Theoretical  Physics,  Kyiv 03143,  Ukraine.}
\affil[3]{Sorbonne Universit\'e, CNRS, Laboratoire de Physique Th\'eorique et Hautes Energies, LPTHE, F-75005 Paris, France.}
\date{}
\begin{document}
\maketitle
\begin{abstract}
  We calculate the three- and four-loop corrections to the massless fermion
propagator in three-dimensional quenched Quantum Electrodynamics with four-component fermions. The three-loop correction is finite and gauge invariant but the four-loop one has singularities except in the Feynman gauge where it is also finite. Our results explicitly show that, up to four loops, gauge-dependent terms are completely determined by lower
order ones in agreement with the Landau-Khalatnikov-Fradkin transformation.
\end{abstract}
\clearpage

\section{Introduction}
\label{sec:Introduction}

Three-dimensional quantum electrodynamics (QED$_3$) is an archetypal gauge field
theory model of strongly interacting relativistic planar fermions. In Euclidean
space, it is described by the action
\begin{equation}
  S=\int \D^3x\left[\frac{1}{4}F^2_{\mu\nu}+\bar\Psi_i\gamma_\mu D_\mu\Psi_i\right],\label{eq:Sdef}
\end{equation}
where $D_\mu=\partial_\mu-ieA_\mu$, $i=1,2,\dots \nf$ where $\nf$ is the number
of four-component massless Dirac fermion flavours, Euclidean gamma matrices
satisfy $\gamma^\dagger_\mu=\gamma_\mu$,
$\{\gamma_\mu,\gamma_\nu\}=\delta_{\mu\nu}$ and the gauge coupling constant
$e^2$ has dimension of mass.

For the past forty years, this super-renormalizable model served as a toy model
for exploring several key problems in quantum field theory such as infrared (IR)
singularities in low-dimensional massless particle
theories~\cite{Jackiw:1980kv,Guendelman1983,Templeton:1981yp,Bergere:1982br,King:1985hr}
(see recent progress in \cite{Karthik:2017hol,Gusynin:2020cra}) and dynamical
symmetry breaking and fermion mass
generation~\cite{Pisarski:1984dj,Appelquist:1986fd,Appelquist:1988sr,Nash:1989xx,Atkinson:1989fp,Pennington:1990bx,Kotikov1993,Gusynin:1995bb,Maris:1996zg,Gusynin:2003ww,Fischer:2004nq}
(see recent progress in
\cite{Gusynin:2016som,Kotikov:2016wrb,Kotikov:2016prf,Karthik:2019mrr,Kotikov:2020slw}).
In the last thirty years, considerable interest in QED$_3$ also came from its
applications to condensed matter physics systems with relativistic-like gapless
quasiparticle excitations at low-energies such as high-$T_c$ superconductors
\cite{Dorey:1991kp,Franz:2001zz,Herbut:2002yq}, planar antiferromagnets
\cite{Farakos:1997hg} and graphene \cite{Semenoff:1984dq} (for graphene, see
reviews in Ref.~\cite{reviews}).

A slight simplification of QED$_3$ takes place in the so-called quenched limit
in which closed fermion loops are set to zero (this amounts to take $\nf=0$).
This limit arose as a useful approximation in the study of the lattice
representation of four-dimensional quantum chromodynamics (QCD$_4$), see
\cite{Parisi1981}, where it was shown that a reasonable estimate of the hadron
spectrum could be obtained by eliminating all internal quark loops. The quenched
limit, and other approximation schemes such as the ladder (rainbow)
approximation, were used in QED$_4$ for a long time to try solving
nonperturbatively a more manageable truncated set of Schwinger-Dyson equations
(see Refs.~\cite{Miransky,Leung1985,Gusynin:1998se} and references therein to
earlier papers). The quenched approximation in QED$_4$, is now still in use in
order to include QED effects in lattice QCD calculations (see the recent
Ref.~\cite{Hatton:2020qhk} and discussions therein).

In the recent paper \cite{Gusynin:2020cra}, we studied the gauge-covariance of
the massless fermion propagator of quenched QED$_3$ in a linear covariant gauge
in dimensional regularization (following
Refs.~\cite{Gusynin:1998se,Kotikov2019,James:2019ctc}). At this point, let's
note that, as in the four-dimensional case, both the fermion propagator and
vertex function of QED$_3$ possess the important property of being covariant
under the Landau-Khalatnikov-Fradkin (LKF) transformations
\cite{Landau1955,Johnson1959}. These transformations, which have a simple form
in coordinate space, allow one to compute Green's functions in an arbitrary
covariant gauge provided they are known in a particular gauge (for applications
of the LKF transformations to QED$_3$, see the papers \cite{Bashir2000} and the
review in Ref.~\cite{Bashir:2007zza}).

The analysis of the LKF transformation of the massless fermion propagator of
quenched QED$_3$ carried out in \cite{Gusynin:2020cra}, led us to reconnect with
the long-standing issue of IR singularities in QED$_3$. In particular, we
concluded that, exactly in three dimensions ($d = 3$), all of the odd
perturbative coefficients of the massless fermion propagator, starting from the
third order one, have to vanish in any gauge in order for quenched QED$_3$ to be
free of (IR) singularities. It turns out that there is a widespread opinion that
quenched perturbation theory is IR finite \cite{Jackiw:1980kv} and recent lattice
studies of quenched QED$_3$ seem to confirm it \cite{Karthik:2017hol}. Usually,
the presence of IR divergences is related to fermion loops. This is because, for
dimensional reasons, higher-order diagrams contain higher powers of momentum in
the denominator. For example, the two-loop fermion self-energy diagram with
vacuum polarization gives rise to a logarithmic divergence~\cite{Jackiw:1980kv} which corresponds to a $1/\ep$-pole in dimensional regularization in $d=3-2\ep$
\cite{Guendelman1983}. However, for the same dimensional reasons, higher-order
quenched diagrams can lead to IR divergences, too. Indeed, it is easy to see
that, at four loops, \eg, the diagram with an insertion of three one-loop
fermion self-energies into a fermion line is logarithmically divergent in gauges
different from the Landau gauge (this was for the first time mentioned in
Ref.~\cite{Templeton:1981yp}). The question then is whether IR divergences of
separate diagrams cancel in their sum or not. The LKF transformation by itself
is unable to provide explicit values for the coefficients in a given gauge.
Order by order calculations are therefore required (at least in a given gauge)
in order to analyze the IR finiteness of QED$_3$ in accordance with our
statement above.

In the present paper, we undertake this task and explicitly calculate the
fermion propagator of quenched QED$_3$ at three and four loops in an arbitrary
linear covariant gauge and in dimensional regularization in $d=3-2\ep$. We find
that the three-loop correction is finite and gauge invariant. Accordingly, the
four-loop one has singularities. Our exact results show that, up to four loops,
gauge-dependent terms are completely determined by lower order ones in perfect
agreement with properties of the LKF transformation following the study
\cite{Gusynin:2020cra}.

The paper is organized as follows. In Sec.~\ref{sec:calcs}, we specify our
notations, provide some details of the calculations and present the results for
the three and four-loop corrections to the fermion propagator. In
Sec.~\ref{sec:LKF3}, we briefly recall the LKF transformation for the fermion
propagator in momentum space and check its consistency with our four-loop
perturbative results. Some predictions beyond four-loops are also presented. In
Sec.~\ref{sec:diags}, a representative sample of the computed diagrams is
presented by focusing on the Landau gauge. The results are summarized and discussed in Sec.~\ref{sec:conclusion}.

\section{Fermion propagator: three- and four-loop coefficients}
\label{sec:calcs}

\subsection{Notations}
\label{sec:calcs:notations}

In the following, we shall consider a Euclidean space of dimension $d=3 - 2\ep$.
The general form of the fermion propagator $S_F(p,\xi)$ in some gauge $\xi$
reads:
\begin{equation}
  \label{eq:SFp}
  S_F(p,\xi) = \frac{\I}{\slashed{p}} \, P(p,\xi) \, , 
\end{equation}
where the tensorial structure, \eg, the factor $\slashed{p}$ containing Dirac
$\gamma$-matrices, has been extracted and $P(p,\xi)$ is a scalar function of
$p=\sqrt{p^2}$.

It is convenient to first express $P(p,\xi)$ as
\begin{equation}
  \label{eq:SFp.1}
  P(p,\xi) = \frac{1}{1-\sigma(p,\xi)} \, ,
\end{equation}
where the 1-particle-irreducible (1PI) part, $\sigma(p,\xi)$, can be represented as
\begin{equation}
  \label{eq:Sigmaxi:QED3}
  \sigma(p,\xi) = \sum_{m=1}^{\infty} \sigma_m(\xi)\, \left(\frac{\al}{2\sqrt{\pi}\,p}\right)^m\,
  {\left(\frac{\bar{\mu}^2}{p^2}\right)}^{m\ep} \, .
\end{equation}
Here, $\sigma_m(\xi)$ are the coefficients of the loop expansion of the fermion
self-energy, $\al=e^2/(4\pi)$ is the dimensionful coupling constant and
$\bar{\mu}$ is the $\overline{\rm{MS}}$-scale.

Following our previous paper \cite{Gusynin:2020cra}, the fermion propagator can
be equivalently represented as
\begin{equation}
  \label{eq:Pxi:QED3}
  P(p,\xi) = \sum_{m=0}^{\infty} a_m(\xi)\, \left(\frac{\al}{2\sqrt{\pi}\,p}\right)^m\,
  {\left(\frac{\bar{\mu}^2}{p^2}\right)}^{m\ep} \, ,
\end{equation}
where $a_m(\xi)$ are now the coefficients of the loop expansion of $P(p,\xi)$.
As will be seen in Sec.~\ref{sec:LKF3}, the form \eqref{eq:Pxi:QED3} is
convenient to study the properties of the propagator under the LKF
transformation.

In both Eqs.~\eqref{eq:Sigmaxi:QED3} and \eqref{eq:Pxi:QED3}, the expansion has
been written in terms of the dimensionless ratio $\al / p$ with an additional
conventional factor of $1/(2\sqrt{\pi})$. Its exact form is coming from the
consideration of four-dimensional QED in \cite{Kotikov2019} (see also
Ref.~\cite{Gusynin:2020cra} and discussions therein). Up to four loops, the
coefficients $a_m(\xi)$ and $\sigma_m(\xi)$ are related to each other as
\begin{equation}
  \label{eq:a.sigma}
  a_1=\sigma_1,\quad
  a_2=\sigma_2+\sigma_1^2,\quad
  a_3=\sigma_3+2\sigma_2\sigma_1+\sigma_1^3,\quad
  a_4=\sigma_4+2\sigma_3\sigma_1+\sigma_2^2+3\sigma_2\sigma_1^2+\sigma_1^4 \, .
\end{equation}

\subsection{Calculational details}
\label{sec:calcs:details}

In quenched QED at 1-, 2-, 3- and 4-loops we encountered 1, 2, 10 and 74 fermion
self-energy diagrams, respectively. Let's note that the two-loop diagrams of QED$_3$ were considered earlier in \cite{Jackiw:1980kv,Guendelman1983}. These papers mainly focused on the IR divergent two-loop diagram (with a fermion loop insertion) which is
absent in the quenched case. In \cite{Guendelman1983}, higher order
diagrams were considered but still with fermion loop insertions. To the
best of our knowledge, the two-loop quenched QED$_3$ fermion propagator
was calculated in \cite{Bashir:2000rv,Bashir:2000iq}. Moreover, all
two-loop diagrams of QED$_3$ were computed in \cite{Kotikov:2013eha} (see
also \cite{Teber:2018jdh}) and their $\ep$-expansion provided near $d=3$. 
The fact that the first singularities in the fermion propagator of QED$_3$ 
without vacuum polarization arise at four loops was mentioned
for the first time in Ref.~\cite{Templeton:1981yp}. However, three- and four-loop corrections to the quenched QED$_3$ fermion propagator have not been previously computed. As will be shown in the next subsections, the three-loop correction is finite but IR singular diagrams do appear at 4-loops in the quenched case (in agreement with \cite{Templeton:1981yp}) and there are 42 of them, the sum of which will be analyzed in the following.

In order to compute all of these diagrams and extract from them the
unrenormalized fermion self-energy of QED$_3$ up to four loops, we first
considered the corresponding results for the unrenormalized QCD quark
propagator. The exact expression for the latter, written in terms of a set of
master integrals and valid for arbitrary space-time dimension $d$ and arbitrary
gauge-fixing parameter $\xi$, is available up to four loops from
\cite{Ruijl:2017eht} and also shipped with the \texttt{FORCER}
package~\cite{Ruijl:2017cxj} designed for the reduction of four-loop massless
propagator-type integrals. The fermion propagator of QED$_d$ is obtained from
this QCD$_d$ result upon performing the following substitutions:
\begin{equation}
  \label{eq:qed2qcdSub}
  C_A = d_A^{abcd}d_A^{abcd} = d_A^{abcd}d_F^{abcd} =0, \quad
  C_F = d_F^{abcd}d_F^{abcd} = T_F = 1\, .
\end{equation}
After that, the quenched limit of QED$_d$ is obtained by setting $\nf = 0$ which
discards all diagrams with closed fermion loops.

The main remaining task was then to compute all required propagator-type master
integrals in an $\ep$-expansion around $d=3$ ($\ep=(3-d)/2$). This could be
achieved with the help of the Dimensional Recurrence and Analyticity (DRA)
method~\cite{Lee:2009dh} which expresses the integrals in the form of fast
convergent sums. The latter are then evaluated with high-precision numerical
values. This in turn allows to reconstruct the analytic expression of master
integrals (in any space-time dimension) with the help of the \texttt{PSLQ}
algorithm~\cite{FBA99} once an adequate basis of transcendental constants is
defined.

We note that near $d=4$ ($\ep=(4-d)/2$), such calculations yield the expansions
of all needed masters~\cite{Lee:2011jt}. The results are well-known and
available in input form for the \texttt{SummerTime} package~\cite{Lee:2015eva}
with the package itself and also from \cite{Magerya:2019cvz}.

The case $d=3-2\varepsilon$ is less well known and was considered in the paper
\cite{Lee:2015eva} from which the $\ep$-expansion of most of the needed master
integrals for the current calculation is available. The successful
reconstructions of \cite{Lee:2015eva} around $d=3$, were carried out using a
basis of transcendental constants consisting only of multiple zeta values (MZV)
and alternating MZVs. As remarked already in \cite{Lee:2015eva}, such a basis is
too restrictive to enable the representation of all of the masters and some of
them were left unreconstructed.

In our work we successfully reconstructed all the needed integrals and found
agreement with results of \cite{Lee:2011jt} using a basis consisting of MZV and
alternating MZVs. On top of that, we encountered one of the constants left
unknown in \cite{Lee:2015eva}. By a careful analysis of the representation of
one such integrals with known closed form expressions in the form of the
${}_3F_2$-functions~\cite{Kotikov:2013eha},\footnote{The results of
\cite{Kotikov:2013eha} were obtained based on the general approach of
Ref.~\cite{Kotikov:1995cw}, where a class of more complicated diagrams with
three arbitrary indices
  was studied and the corresponding results were expressed in terms of
combinations of ${}_3F_2$-hypergeometric functions of unit argument.} we found
that elements of its $\varepsilon$-expansion belong to the set of generalized
polylogarithms (GPLs) with fourth-root of unity alphabet. Extending our
\texttt{PSLQ} basis to include the full set of GPLs with fourth root of unity
arguments we successfully reconstructed its analytical value
\tikzset{sl/.style={line width=1pt,draw=bla}}
\begin{equation}
  \label{eq:M22res}
  \vcenter{\hbox{
      \begin{tikzpicture}[use Hobby shortcut, scale=0.8]
        \draw[sl] (0,0) circle (1);
        \coordinate (vL) at (180:1);
        \coordinate (vR) at (0:1);
        \coordinate (vU) at (90:1);
        \coordinate (vD) at (270:1);
        \coordinate (v1) at (45:0.5);
        \coordinate (v2) at (225:0.5);
        \draw[sl] (vU) .. (v1) .. (vR);
        \draw[sl] (vL) .. (v2) .. (vD);
        \draw[sl] (vU) -- (vD);
        \draw[sl] (vL) -- (180:1.2);
        \draw[sl] (vR) -- (0:1.2);
        \fill (vL) circle (1pt);
        \fill (vR) circle (1pt);
        \fill (vU) circle (1pt);
        \fill (vD) circle (1pt);
      \end{tikzpicture}
    }}
  =
    \vcenter{\hbox{
      \begin{tikzpicture}[use Hobby shortcut, scale=0.8]
        \draw[sl] (0,0) circle (1);
        \coordinate (vL) at (180:1);
        \coordinate (vR) at (0:1);
        \coordinate (v1) at (90:0.5);
        \coordinate (v2) at (270:0.5);
        \draw[sl] (vL) .. (v1) .. (vR);
        \draw[sl] (vL) .. (v2) .. (vR);
        \draw[sl] (vL) -- (vR);
        \draw[sl] (vL) -- (180:1.2);
        \draw[sl] (vR) -- (0:1.2);
        \fill (vL) circle (1pt);
        \fill (vR) circle (1pt);
      \end{tikzpicture}
    }}
  \cdot
  \left(
    -8 \varepsilon \left(\mathcal{C} \pi^2 + 24 \mathrm{Cl}_4\left(\frac{\pi}{2}\right)\right)
    + O(\varepsilon^2)
  \right)\, ,
\end{equation}
where we factored out the four-loop sunset integral to follow the normalization
prescriptions of \cite{Lee:2015eva}. In Eq.~\eqref{eq:M22res}, $\mathcal{C}={\rm
Cl}_2(\pi/2)$ is Catalan's constant and $\mathrm{Cl_n(\theta)}$ is Clausen's
function which, for even weight, can be expressed through the classical
polylogarithm as $\mathrm{Cl}_{2k}(\theta) = \mathrm{Im}\
\mathrm{Li}_{2k}\left(e^{i\theta}\right)$. As can be understood from the above
result, the required extension of the basis of transcendental constants includes
polylogarithms with fourth-root of unity argument, in the present case Clausen's
function (see, for example, Ref.~\cite{Laporta:2018eos}, where GPLs with
second-, fourth- and sixth-root of unity arguments appear).

\subsection{\texorpdfstring{Results for the fermion self-energy \eqref{eq:Sigmaxi:QED3}}{Results for the fermion self-energy}}
\label{sec:calcs:res:sigma}

We now present our results for $\sigma_m(\xi)$ which can be represented as
\begin{equation}
  \label{eq:sim=sim0}
  \sigma_m(\xi)= \sigma_m(0) + \,\xi \, \tilde{\sigma}_m(\xi) \, ,
\end{equation}
where we have explicitly separated the part independent 
from $\xi$ which corresponds to the full result in the Landau gauge.

Considering the first two orders of the $\ep$-expansion, we have for the coefficients $\sigma_m(0)$
\begin{subequations}
  \label{eq:coeffssigma0}
  \begin{align}
    \sigma_1(0) & = 0\,; 
    \label{eq:a1.0} \\
    \sigma_2(0) & = \pi \left[ \frac{3 \pi^2}{4} - 7 - \Bigl((1-3 \, l_2)\pi^2 + 12 \Bigr)\ep \right]\,;
                  \label{eq:a12.0} \\
    \sigma_3(0) & = \pi^{5/2} \left[ \frac{43\pi^2}{4} - 105 + \ep \left\{ 2 \Bigl(185 - 105 \, l_2 + 137\zeta_3\Bigr)
                  -\frac{\pi^2}{6} \Bigl( 451 - 171 \, l_2 \Bigr)\right\}\right]\,;
                  \label{eq:a3.0} \\
    \sigma_4(0) & =
                  \pi^2 \left[ \left(\frac{43}{6}\pi^2-70\right) \, \frac{1}{\ep} +
                  \bar{\sigma}_{4} + \frac{5954}{3} + \frac{173}{18}\,\pi^2- \frac{513}{10}\,\pi^4\right]\, ,
                  \label{eq:si4.0}
  \end{align}
\end{subequations}
where
$\bar{\sigma}_{4}$ contains the most complicated part
\begin{equation}
  \label{eq:oa40}
  \bar{\sigma}_{4} = 209 \, l_2^4 + 5016 \, a_4
  + 4264 \, \mathrm{Cl}_4(\pi/2) + \left(\frac{533}{3} \mathcal{C} -930\,l_2\right) \, \pi^2
  +\frac{2078}{3} \zeta_3 \, ,
\end{equation}
and
\begin{equation}
  \label{eq:l2}
  l_2 = \ln 2,\quad
  a_4 = \mathrm{Li}_4(1/2),\quad
  \zeta_n = \mathrm{Li}_n(1) \, ,
\end{equation}
where ${\rm Li}_n$ are polylogarithms.

With the same accuracy, we have for the coefficients $\tilde{\sigma}_m(\xi)$
\begin{subequations}
  \label{eq:coeffssigmaxi}
  \begin{align}
    \tilde{\sigma}_1(\xi) & = - \frac{\pi^{3/2}}{2} \Bigl(1 - 2 (1 - l_2)\ep\Bigr)\,;
    \\
    \tilde{\sigma}_2(\xi) & = \pi \, \xi \left[1-\frac{\pi^2}{4} -\bigl(4-(1 - l_2) \pi^2\bigr)\ep\right]\,;
    \\
    \tilde{\sigma}_3(\xi) & = \pi^{5/2}\, \Biggl[\frac{3\pi^2}{4} - 7
                            + \left(1 - \frac{\pi^2}{8}\right)\xi^2
                            + \ep \Biggl\{-40-14l_2+ \frac{\pi^2}{2} \bigl(4+9l_2\bigr) \nonumber \\
                          & \hspace{1cm}
                            + \left(2 \, l_2 - 4 + \frac{3\pi^2}{4} \bigl(1 - l_2\bigr)
                            \right) \xi^2 \Biggr\}
                            \Biggr] \,; 
    \\
    \tilde{\sigma}_4(\xi) & = \pi^{2} \Biggl[\left( 70
                            - \frac{43\pi^2}{6} \right)\, \frac{1}{\ep}
                            + \frac{520}{3} - \frac{\pi^2}{9} \Bigl(881 + 42l_2\Bigr)
                            + \frac{129\pi^4}{27}
                            -\frac{548}{3} \zeta_3  \nonumber \\
                          & \hspace{1cm}
                            + \xi \, \left(28-\frac{33\pi^2}{4} + \frac{9\pi^4}{16} \right)
                            + \xi^3 \, \left(-\frac{4}{3} +\frac{3\pi^2}{4}
                            - \frac{\pi^4}{16} \right)\Biggr]\, .
                            \label{eq:tsi4xi}
  \end{align}
\end{subequations}

We would like to note that the finite parts of the coefficients $\sigma_1(\xi)$
and $\sigma_2(\xi)$ coincide with the corresponding ones in
Ref.~\cite{Kotikov:2013eha}.

Moreover, from Eqs.~\eqref{eq:si4.0} and \eqref{eq:tsi4xi}, we notice that
\begin{equation}
  \label{eq:si4.xi}
  \sigma_{4}(\xi) = \pi^2 \left(\frac{43}{6}\pi^2 - 70\right) \,
  \frac{(1 - \xi)}{\ep} + O(\ep^0) \, ,
\end{equation}
\ie, the total four-loop contribution is finite in the Feynman gauge.

\subsection{\texorpdfstring{Results for the fermion propagator \eqref{eq:Pxi:QED3}}{Results for the fermion propagator}}
\label{sec:calcs:res:propagator}

As in the case of $\sigma_m(\xi)$ in \eqref{eq:sim=sim0}, it is convenient to
present the results for $a_m(\xi)$ in the form
\begin{equation}
  \label{eq:am=am0}
  a_m(\xi)= a_m(0) + \,\xi \, \tilde{a}_m(\xi)\, ,
\end{equation}
where we have also explicitly separated the part independent 
from $\xi$ which corresponds to the full result in the Landau gauge.

Since $\sigma_1(\xi) \sim \xi$, we see from \eqref{eq:a.sigma} that
$a_i(0)=\sigma_i(0)$ for $i \leq 3$ and thus $a_i(0)$ with $i \leq 3$ can
straightforwardly be read off from Eqs.~\eqref{eq:a1.0}, \eqref{eq:a12.0} and \eqref{eq:a3.0}. In
agreement with \eqref{eq:a.sigma}, we have for $a_{4}(0)$
\begin{equation}
  \label{eq:a4.0}
  a_{4}(0) = \sigma_4(0)
  + \pi^2 {\left(\frac{3\pi^2}{4} - 7\right)}^2
  = \pi^2 \left[ \left(\frac{43}{6}\pi^2 - 70\right) \, \frac{1}{\ep} + \bar{\sigma}_{4}
    + \frac{6101}{3} - \frac{8}{9}\,\pi^2 - \frac{4059}{80}\,\pi^4\right]\, ,
\end{equation}
where $\bar{\sigma}_{4}$ was defined in Eq.~\eqref{eq:oa40}.

With the same accuracy, we have for the coefficients  $\tilde{a}_m(\xi)$
\begin{subequations}
  \label{eq:coeffa0}
  \begin{align}
    \tilde{a}_1(\xi) & =\tilde{\sigma}_1(\xi)=-\frac{\pi^{3/2}}{2} \Bigl(1-2(1-l_2)\ep\Bigr)\,;
    \\
    \tilde{a}_2(\xi) & =\pi \, \xi \Bigl(1-4\ep\Bigr)\,;
    \\
    \tilde{a}_3(\xi) & = \pi^{5/2}\,\ep \left(\frac{43\pi^2}{4}-105 +2\xi^2\right)\,;
                       \label{eq:ta3xi}\\
    \tilde{a}_4(\xi) & = \frac{\pi^{2}}{3}\left[\left(210-\frac{43\pi^2}{2} \right)\, \frac{1}{\ep}
                       + 520 + \frac{2\pi^2}{3}\Bigl(32 - 21 \, l_2\Bigr) - 548\zeta_3
                       + 6\xi \, \left(7 - \frac{3\pi^2}{4}\right) - \xi^3\right]
                       \label{eq:ta4.xi}\, .
  \end{align}
\end{subequations}

We would like to note that the finite parts of the coefficients $a_1(\xi)$ and
$a_2(\xi)$ coincide with the corresponding ones in Ref.~\cite{Bashir2000}
(see also Ref.~\cite{Gusynin:2020cra} and discussions therein).

From the above results, we see that the coefficients $\tilde{a}_m(\xi)$
$(m=2,3,4)$ have simpler forms than the corresponding coefficients
$\tilde{\sigma}_m(\xi)$. Moreover, as in the case of $\sigma_{4}(\xi)$, we
notice from Eqs.~\eqref{eq:a4.0} and \eqref{eq:ta4.xi} that
\begin{equation}
  \label{eq:a4.xi}
  a_{4}(\xi) = \sigma_{4}(\xi)  + O(\ep^0)
  = \frac{2 \pi^2}{3}\, \left(\frac{43\pi^2}{4} - 105 \right) \,
  (1-\xi) \, \frac{1}{\ep} + O(\ep^0) \, ,
\end{equation}
\ie, the total four-loop contribution is finite in the Feynman gauge.

\section{LKF transformation}
\label{sec:LKF3}

\subsection{Comparison with the perturbative results up to four loops}

It is convenient to introduce the $x$-space representation $S_F(x,\xi)$ of the
fermion propagator as:
\begin{equation}
  \label{eq:SFx}
  S_F(x,\xi) =  \slashed{x} \, X(x,\xi) \,
\end{equation}
which is related by the Fourier transform to $S_F(p,\xi)$ in \eqref{eq:SFp}. The
LKF transformation expresses the covariance of the fermion propagator under a
gauge transformation. It can be derived by standard arguments, see, \eg,
\cite{Landau1955,Johnson1959} and its general form can be written as (see
Refs.~\cite{Kotikov2019,James:2019ctc}):
\begin{equation}
  \label{eq:def:D(x):QED3}
  S_F(x,\xi) = S_F(x,\eta)\,e^{D(x)}\, ,\quad
  D(x) = e^2\,\Delta\,\mu^{2\ep}\,\int \frac{\D^d q}{(2\pi)^d } \,
  \frac{e^{- i q x}}{q^4},\quad
  \Delta=\xi -\eta \, ,
\end{equation}
in $d=3-2\ep$. The calculation~\cite{Gusynin:1998se} yields:
\begin{equation}
  \label{eq:DxQED3}
  D(x) = - \frac{\al\,\Delta}{2\pi\,\mu}\,\frac{\Gamma(1/2-\ep)}{1+2\ep} \,
  (\pi \mu^2 x^2)^{1/2+\ep} \, .
\end{equation}

The LKF transformation \eqref{eq:def:D(x):QED3} relates~\cite{Gusynin:2020cra}
the coefficients $a_k(\xi)$ and $a_m(\xi)$ in \eqref{eq:Pxi:QED3} as
\begin{equation}
  \label{eq:am.xid3}
  a_k(\xi) = \sum_{m=0}^{k}
  \, (-2 \Delta)^{k-m} \, a_m(\eta) \,
  \Phi(m,k,\ep) \  \phi(k-m,\ep),
\end{equation}
where
\begin{equation}
  \label{eq:hPhi}
  \Phi(m,k,\ep)
  = \frac{\Gamma(3/2-m/2-(m+1)\ep)\Gamma(1+k/2+k\ep)}
  {\Gamma(1+m/2+m\ep)\Gamma(3/2-k/2-(k+1)\ep)}
\end{equation}
and
\begin{equation}
  \label{eq:phi}
  \phi(l,\ep) = \frac{\Gamma^l(1/2-\ep)}{l!\,(1+2\ep)^l\Gamma^l(1+\ep)}\, .
\end{equation}

Consider $a_{m}(\xi)$ with $m\leq 4$. Keeping only the first two orders of the
$\ep$-expansion, we have:
\begin{subequations}
  \label{eq:axi16}
  \begin{align}
    a_{0}(\xi) & = a_{0}(\eta) =1 \, ,
                 \label{eq:sc:axi0} \\
    a_{1}(\xi) & = a_{1}(\eta) - \frac{\pi}{2} \, \delta \,
                 \Bigl(1+2\ep(l_2-1)\Bigr) \, a_{0}(\eta)\, ,
                 \label{sc:axi1} \\
    a_{2}(\xi) & = a_{2}(\eta) - \frac{4}{\pi} \, \delta \,
                 \Bigl(1-2\ep(l_2+1)\Bigr) \, a_{1}(\eta)
                 + \delta^2 \,\Bigl(1-4\ep\Bigr)
                 \, a_{0}(\eta)\, ,
                 \label{sc:axi2} \\
    a_{3}(\xi) & = a_{3}(\eta)  + 6\pi\ep \, \delta \, a_{2}(\eta)
                 - 12\ep \, \delta^2 \, a_{1}(\eta)
                 +2\pi\ep \, \delta^3 \, a_{0}(\eta)\, ,
                 \label{sc:axi3} \\
    a_{4}(\xi) & = a_{4}(\eta) - \frac{2\delta}{3\pi\ep} \,
                 \Bigl(1+2\ep(3-l_2)\Bigr)\, a_{3}(\eta) - 2 \delta^2 \, a_{2}(\eta)
                 + \frac{8\delta^3 }{3\pi} \, a_{1}(\eta)
                 -\frac{\delta^4 }{3} \, a_{0}(\eta)\, ,
                 \label{sc:axi4}
  \end{align}
\end{subequations}
where $\delta=\sqrt{\pi}\Delta$.

Setting $\eta=0$, \ie, choosing the initial gauge as the Landau gauge, we can
see that our results for $\tilde{a}_{m}(\xi)$ are completely determined by
$a_{l}(\xi)$, $(l < m)$, \ie, by the coefficients of lower orders in agreement
with the properties of the LKF transformation.

Moreover, the results of Eqs.~\eqref{eq:axi16} are in full agreement with the
perturbative results presented in Sec.~\ref{sec:calcs:res:propagator}.

\subsection{Beyond four-loops}
\label{sec:LKF3:bfl}

As can be seen from \eqref{eq:a4.xi}, the singularity of the four-loop
coefficient $a_{4}(\xi)$ is $\sim (1-\xi)$, \ie, the fermion propagator
including up to four-loop corrections is finite in the Feynman gauge. This
intriguing fact calls for a closer examination of higher order contributions
and, as a first try, we will proceed by using the LKF transformation.

We therefore consider $a_{5}(\xi)$ and $a_{6}(\xi)$. From Eq.~\eqref{eq:am.xid3}, we have: 
\begin{subequations}
  \label{eq:axi56}
  \begin{align}
    a_{5}(\xi) & = a_{5}(\eta)  + \frac{45}{2}\pi\ep \, \delta \,
                 a_{4}(\eta) - \frac{15}{2} \, \delta^2 \, a_{3}(\eta)
                 - 15\pi \ep \, \delta^3 \, a_{2}(\eta) + 15\ep \, \delta^4 \, a_{1}(\eta)
                   - \frac{3}{2} \pi\ep \, \delta^5 \, a_{0}(\eta)\, ,
                 \label{eq:sc:axi5} \\
    a_{6}(\xi)  & = a_{6}(\eta) + \frac{4\delta}{5\pi\ep} \,
                  a_{5}(\eta) - 9 \delta^2 \, a_{4}(\eta)
                  + \frac{2\delta^3 }{\pi\ep} \, a_{3}(\eta)
                 + 3 \delta^4 \, a_{2}(\eta)
                 - \frac{12\delta^5 }{5\pi} \, a_{1}(\eta)
                 + \frac{\delta^6 }{5} \, a_{0}(\eta)\, .
                 \label{eq:sc:axi6}
  \end{align}
\end{subequations}
We may then take the $\eta$-gauge as the Feynman gauge and consider $a_{5}(\xi)$
and $a_{6}(\xi)$ with accuracies $O(\ep)$ and $O(\ep^0)$, respectively. This
yields:
\begin{subequations}
  \label{eq:axi56a}
  \begin{align}
    a_{5}(\xi) & =a_{5}(1)  - \frac{15}{2} \, \pi \, (\xi-1)^2 \, a_{3} + O(\ep)\, ,
                 \label{eq:sc:axi5a} \\
    a_{6}(\xi) & = a_{6}(1) + \frac{4(\xi-1)}{5\sqrt{\pi}\ep} \, a_{5}(1)
                 + \frac{2\sqrt{\pi}(\xi-1)^3 }{\ep} \, a_{3} + O(\ep^0)\, ,
                 \label{eq:sc:axi6a}
  \end{align}
\end{subequations}
where we took into account the fact that the finite part of $a_{3}$ is
gauge-independent.

From these results, we see that the LKF transformation gives information about
the $\xi$-dependence of $a_{5}(\xi)$ and $a_{6}(\xi)$, as expected. Some
singularities may still be hidden in $a_{6}(1)$ and further understanding of the
singular structure of $a_{6}(\xi)$ requires explicit $5$- and $6$-loop
computations (at least in a specific gauge).

\section{Diagrams in the Landau gauge}
\label{sec:diags}

As we discussed in Sec.~\ref{sec:calcs:details}, there is a total of 87 diagrams
to compute in order to derive the fermion propagator of quenched QED$_3$ up to
4-loops with an arbitrary gauge-fixing parameter $\xi$. The results presented in
Secs.~\ref{sec:calcs:res:sigma} and \ref{sec:calcs:res:propagator}, were
obtained by computing all of these diagrams.

In order to provide the interested reader with a representative sample of the
graphs, we focus in this section on the Landau gauge. The reason is that it is
the gauge where there is the least number of diagrams as most of them vanish in
the limit $\xi=0$. Moreover, as discussed in Sec.~\ref{sec:LKF3}, it is enough
to compute the fermion propagator in this gauge as the LKF transformation allows
to reconstruct the full $\xi$-dependence of the propagator.

In the Landau gauge, there is no one-loop contribution and there are 1, 4 and 30
diagrams at two-, three- and four-loops, respectively; so a total of 35
diagrams. Focusing on the leading order contribution to the $\ep$-expansion of
these diagrams, the two- and three-loop contributions will be considered with an
accuracy $O(\ep)$ and the four-loop contributions with an accuracy $O(\ep^0)$.
So, amongst the 30 four-loop diagrams only the 8 divergent ones need to be
considered (the other 22 diagrams are finite). Moreover, taking into account the
fact that mirror conjugate graphs take the same value, we are left with only 3
distinct graphs at 3-loops and 4 distinct graphs at 4-loops. Hence a total of 8
distinct diagrams contribute to the Landau gauge quenched QED$_3$ fermion
propagator up to 4 loops.

For the sake of clarity, we explicitly display the distinct graphs together with
their values. The single diagram contributing at two-loop level is given by:
\tikzset{
  di/.style={line width=1pt,draw=black, postaction={decorate},
    decoration={markings,mark=at position .55 with
      {\arrow[scale=1,draw=black,>=stealth]{>}}}},
  photon/.style={draw=bla,line width=1pt,decorate,
    decoration={snake,aspect=0.5,amplitude=1.5pt,segment
      length=5pt}}}
\begin{equation}
  \label{eq:dia2l}
  \vcenter{\hbox{
  \begin{tikzpicture}[scale=1]
    \coordinate (vi) at (-1.8,0);
    \coordinate (vo) at (1.8,0);    
    \coordinate (v180) at (180:1);
    \coordinate (v90) at (90:1);
    \coordinate (v270) at (270:1);
    \coordinate (v0) at (0:1);    
    \draw [di] (vi) -- (v180);
    \draw [di] (v180) arc (180:270:1);
    \draw [di] (v270) -- (v90);
    \draw [di] (v90) arc (90:0:1);
    \draw [photon] (v90) arc (90:180:1);
    \draw [photon] (v270) arc (270:360:1);
    \draw [di] (v0) -- (vo);
    \fill (v0) circle (2pt);
    \fill (v90) circle (2pt);
    \fill (v180) circle (2pt);
    \fill (v270) circle (2pt);
  \end{tikzpicture}
  }}
  = -\frac{1}{4}\pi \left(28-3\pi^2\right) + O(\varepsilon) \, .
\end{equation}
%
%
At three loops, the two benz diagrams in \eqref{eq:mconj3l} are mirror conjugate
to eachother and therefore share the same value. Hence, the three distinct
three-loop graphs read:
\begin{subequations}
\label{tab:3loopgraphs}
\begin{align}
  \vcenter{\hbox{
  \begin{tikzpicture}[scale=1]
    \coordinate (vi) at (-1.8,0);
    \coordinate (vo) at (1.8,0);    
    \coordinate (v180) at (180:1);
    \coordinate (v90) at (90:1);
    \coordinate (v270) at (270:1);
    \coordinate (v0) at (0:1);
    \coordinate (v60) at (60:1);    
    \coordinate (v120) at (120:1);
    \coordinate (v240) at (240:1);
    \coordinate (v300) at (300:1);    
    \draw [di] (vi) -- (v180);
    \draw [di] (v0) -- (vo);
    \draw [di] (v180) arc (180:120:1);
    \draw [di] (v300) arc (300:360:1);
    \draw [di] (v120) arc (120:60:1);
    \draw [di] (v240) arc (240:300:1);
    \draw [photon] (v180) arc (180:240:1);
    \draw [photon] (v0) arc (0:60:1);
    \draw [di] (v60) -- (v240);
    \draw [photon] (v120) -- (120:0.2);
    \draw [photon] (300:0.2) -- (v300);
    \fill (v0) circle (1.5pt);
    \fill (v60) circle (1.5pt);
    \fill (v120) circle (1.5pt);
    \fill (v180) circle (1.5pt);
    \fill (v240) circle (1.5pt);
    \fill (v300) circle (1.5pt);
  \end{tikzpicture}
  }}
  & = \frac{3}{2}\pi^{5/2} \left(10 - \pi^2\right) + O(\varepsilon) \, ,
  \\
  \vcenter{\hbox{
  \begin{tikzpicture}[scale=1]
    \coordinate (vi) at (-1.8,0);
    \coordinate (vo) at (1.8,0);    
    \coordinate (v180) at (180:1);
    \coordinate (v90) at (90:1);
    \coordinate (v270) at (270:1);
    \coordinate (v0) at (0:1);
    \coordinate (v60) at (60:1);    
    \coordinate (v120) at (120:1);
    \coordinate (v240) at (240:1);
    \coordinate (v300) at (300:1);    
    \draw [di] (vi) -- (v180);
    \draw [di] (v0) -- (vo);
    \draw [di] (v180) arc (180:240:1);
    \draw [di] (v300) arc (300:360:1);
    \draw [di] (v120) arc (120:60:1);
    \draw [photon] (v240) arc (240:300:1);
    \draw [photon] (v180) arc (180:120:1);
    \draw [photon] (v0) arc (0:60:1);
    \draw [di] (v240) -- (v120);
    \draw [di] (v60) -- (v300);
    \fill (v0) circle (1.5pt);
    \fill (v60) circle (1.5pt);
    \fill (v120) circle (1.5pt);
    \fill (v180) circle (1.5pt);
    \fill (v240) circle (1.5pt);
    \fill (v300) circle (1.5pt);
  \end{tikzpicture}
  }}
  & = - \frac{1}{12}\pi^{5/2} \left(480-49\pi^2\right) + O(\varepsilon) \, ,
  \\
  2 \ast \vcenter{\hbox{
  \begin{tikzpicture}[scale=1]
    \coordinate (vi) at (-1.8,0);
    \coordinate (vo) at (1.8,0);    
    \coordinate (v180) at (180:1);
    \coordinate (v90) at (90:1);
    \coordinate (v270) at (270:1);
    \coordinate (v0) at (0:1);
    \coordinate (v60) at (60:1);    
    \coordinate (v120) at (120:1);
    \coordinate (v240) at (240:1);
    \coordinate (v300) at (300:1);
    \coordinate (vc) at (0,0);    
    \draw [di] (vi) -- (v180);
    \draw [di] (v0) -- (vo);
    \draw [di] (v180) arc (180:120:1);
    \draw [di] (v120) arc (120:60:1);
    \draw [photon] (v0) arc (0:60:1);
    \draw [photon] (v180) arc (180:270:1);
    \draw [di] (v270) arc (270:360:1);
    \draw [di] (v60) -- (vc);
    \draw [di] (vc) -- (v270);
    \draw [photon] (v120) -- (vc);
    \fill (v0) circle (1.5pt);
    \fill (v60) circle (1.5pt);
    \fill (v120) circle (1.5pt);
    \fill (v180) circle (1.5pt);
    \fill (v270) circle (1.5pt);
    \fill (vc) circle (1.5pt);
  \end{tikzpicture}
  }}
  & = - \frac{1}{6}\pi^{5/2} \left(480-49\pi^2\right) + O(\varepsilon) \, .
    \label{eq:mconj3l}
\end{align}
\end{subequations}
Similarly, the four-loop diagrams are grouped in pairs of mirror conjugate
graphs and the four leading contributions are given by:
\begin{subequations}
\label{tab:4loopgraphs}
\begin{align}
  2 \ast \vcenter{\hbox{
  \begin{tikzpicture}[scale=1]
    \coordinate (vi) at (-1.8,0);
     \coordinate (vo) at (1.8,0);    
    \coordinate (v180) at (180:1);
    \coordinate (v90) at (90:1);
    \coordinate (v270) at (270:1);
    \coordinate (v0) at (0:1);
    \coordinate (v60) at (60:1);    
    \coordinate (v120) at (120:1);
    \coordinate (v240) at (240:1);
    \coordinate (v300) at (300:1);
    \coordinate (v225) at (225:1);
    \coordinate (v315) at (315:1);
    \coordinate (vc1) at (0.2,0.5);
    \coordinate (vc2) at (0.5,-0.2);    
    \draw [di] (vi) -- (v180);
    \draw [di] (v0) -- (vo);
    \draw [di] (v180) arc (180:225:1);
    \draw [di] (v225) arc (225:270:1);
    \draw [di] (v270) arc (270:315:1);
    \draw [photon] (v315) arc (315:360:1);
    \draw [photon] (v180) arc (180:90:1);
    \draw [di] (v90) arc (90:0:1);
    \draw [photon] (v270) -- ($(v270)!0.35!(vc1)$);
    \draw [photon] (vc1) -- ($(vc1)!0.45!(v270)$);
    \draw [photon] (v225) -- (vc2);
    \draw [di] (v315) -- (vc2);
    \draw [di] (vc2) -- (vc1);
    \draw [di] (vc1) -- (v90);
    \fill (v0) circle (1.5pt);
    \fill (v90) circle (1.5pt);
    \fill (v180) circle (1.5pt);
    \fill (v225) circle (1.5pt);
    \fill (v270) circle (1.5pt);
    \fill (v315) circle (1.5pt);
    \fill (vc1) circle (1.5pt);
    \fill (vc2) circle (1.5pt);
  \end{tikzpicture}
  }}
  & = \frac{1}{\ep}\pi^{2}(10-\pi^2) + O(1)
  \\
  2 \ast \vcenter{\hbox{
  \begin{tikzpicture}[scale=1]
    \coordinate (vi) at (-1.8,0);
    \coordinate (vo) at (1.8,0);    
    \coordinate (v180) at (180:1);
    \coordinate (v270) at (270:1);
    \coordinate (v0) at (0:1);
    \coordinate (v60) at (60:1);    
    \coordinate (v120) at (120:1);
    \coordinate (vc1) at (-0.5,0.3);
    \coordinate (vc2) at (0.5,0.3);
    \coordinate (vc3) at (0,-0.4);    
    \draw [di] (vi) -- (v180);
    \draw [di] (v0) -- (vo);
    \draw [di] (v180) arc (180:120:1);
    \draw [di] (v120) arc (120:60:1);
    \draw [photon] (v60) arc (60:0:1);
    \draw [photon] (v180) arc (180:270:1);
    \draw [di] (v270) arc (270:360:1);
    \draw [photon] (v120) -- (vc1);
    \draw [photon] (vc3) -- (vc2);
    \draw [di] (vc3) -- (v270);
    \draw [di] (vc1) -- (vc3);
    \draw [di] (vc2) -- (vc1);
    \draw [di] (v60) -- (vc2);
    \fill (v0) circle (1.5pt);
    \fill (v60) circle (1.5pt);
    \fill (v120) circle (1.5pt);
    \fill (v180) circle (1.5pt);
    \fill (v270) circle (1.5pt);
    \fill (vc3) circle (1.5pt);
    \fill (vc1) circle (1.5pt);
    \fill (vc2) circle (1.5pt);
  \end{tikzpicture}
  }}
  & = - \frac{1}{18 \ep}\pi^{2}(480 - 49 \pi^2) + O(1)
  \\
  2 \ast \vcenter{\hbox{
  \begin{tikzpicture}[scale=1]
    \coordinate (vi) at (-1.8,0);
    \coordinate (vo) at (1.8,0);    
    \coordinate (v180) at (180:1);
    \coordinate (v270) at (270:1);
    \coordinate (v0) at (0:1);
    \coordinate (v45) at (45:1);    
    \coordinate (v90) at (90:1);
    \coordinate (v135) at (135:1);
    \coordinate (vc1) at (110:0.5);
    \coordinate (vc2) at (-45:0.3);
    \draw [di] (vi) -- (v180);
    \draw [di] (v0) -- (vo);
    \draw [di] (v180) arc (180:135:1);
    \draw [photon] (v135) arc (135:90:1);
    \draw [di] (v90) arc (90:45:1);
    \draw [photon] (v45) arc (45:0:1);
    \draw [photon] (v180) arc (180:270:1);
    \draw [di] (v270) arc (270:360:1);
    \draw [di] (vc2) -- (v270);
    \draw [di] (v45) -- (vc2);
    \draw [di] (v135) -- (vc1);
    \draw [di] (vc1) -- (v90);
    \draw [photon] (vc1) -- (vc2);
    \fill (v0) circle (1.5pt);
    \fill (v45) circle (1.5pt);
    \fill (v90) circle (1.5pt);
    \fill (v135) circle (1.5pt);
    \fill (v180) circle (1.5pt);
    \fill (v270) circle (1.5pt);
    \fill (vc1) circle (1.5pt);
    \fill (vc2) circle (1.5pt);
  \end{tikzpicture}
  }}
  & = - \frac{1}{18 \ep}\pi^{2}(480 - 49 \pi^2) + O(1)
  \\
  2 \ast \vcenter{\hbox{
  \begin{tikzpicture}[scale=1]
    \coordinate (vi) at (-1.8,0);
    \coordinate (vo) at (1.8,0);    
    \coordinate (v180) at (180:1);
    \coordinate (v270) at (270:1);
    \coordinate (v0) at (0:1);
    \coordinate (v45) at (45:1);    
    \coordinate (v90) at (90:1);
    \coordinate (v135) at (135:1);
    \coordinate (vc1) at (110:0.5);
    \coordinate (vc2) at (-45:0.3);
    \draw [di] (vi) -- (v180);
    \draw [di] (v0) -- (vo);
    \draw [photon] (v180) arc (180:135:1);
    \draw [di] (v135) arc (135:90:1);
    \draw [di] (v90) arc (90:45:1);
    \draw [di] (v45) arc (45:0:1);
    \draw [di] (v180) arc (180:270:1);
    \draw [photon] (v270) arc (270:360:1);
    \draw [di] (v270) -- (vc2);
    \draw [photon] (v45) -- (vc2);
    \draw [di] (vc1) -- (v135);
    \draw [photon] (vc1) -- (v90);
    \draw [di] (vc2) -- (vc1);
    \fill (v0) circle (1.5pt);
    \fill (v45) circle (1.5pt);
    \fill (v90) circle (1.5pt);
    \fill (v135) circle (1.5pt);
    \fill (v180) circle (1.5pt);
    \fill (v270) circle (1.5pt);
    \fill (vc1) circle (1.5pt);
    \fill (vc2) circle (1.5pt);
  \end{tikzpicture}
  }}
  & = - \frac{1}{18 \ep}\pi^{2}(480 - 49 \pi^2) + O(1)
\end{align}
\end{subequations}

Summing all of the above contributions order by order in the loop expansion
yields the coefficients $\sigma_i(0)$ ($i=1-4$) in agreement with
Eqs.~(\ref{eq:coeffssigma0}) at the leading order of the $\ep$-expansion. With
the accuracy used, these coefficients are equal to the coefficients $a_i(0)$
($i=1-4$). Substituting them in Eqs.~(\ref{eq:axi16}) with $\eta=0$ allows to
reconstruct the gauge-dependent part of $a_i(\xi)$ ($i=1-4$) in agreement with
Eqs.~(\ref{eq:coeffa0}) at the leading order of the $\ep$-expansion.

\section{Summary and Conclusion}
\label{sec:conclusion}

In the present paper, we have examined the perturbative structure of the
massless fermion propagator of quenched QED$_3$ up to four loops.

Our study was motivated by our recent publication \cite{Gusynin:2020cra} where
the gauge covariance of the fermion propagator of quenched QED$_3$ was studied
using the LKF transformation in dimensional regularization ($d=3-2\ep$). This
non-perturbative transformation revealed an interesting parity effect, whereby
the contributions of odd orders, starting from the third one, to even orders are
accompanied by singularities taking the form of poles, $\ep^{-1}$, in
dimensional regularization. In turn, even orders produce contributions to odd
ones, starting from the third order, which are $\sim \ep$.

Following arguments in favor of the IR (and ultraviolet) perturbative finiteness
of massless quenched
QED$_3$~\cite{Jackiw:1980kv,Karthik:2017hol} and therefore
assuming the existence of a finite limit as $\ep \to 0$, we concluded in
Ref.~\cite{Gusynin:2020cra} that, exactly in $d=3$, all odd coefficients
$a_{2t+1}(\xi)$ in perturbation theory, except $a_{1}$, should be exactly zero
in any gauge.

This statement needed a check since analytical expressions for
the fermion self-energy diagrams were known only at two-loop order. 
This is what we have done in the present paper by computing the three- and four-loop corrections to the
massless fermion propagator, \ie, the coefficients $a_{3}(\xi)$ and
$a_{4}(\xi)$, directly in the framework of perturbation theory (see
Sec.~\ref{sec:diags} for some details on the computed diagrams in the Landau
gauge). We found that $a_{3}(\xi)$ is finite and gauge-independent when $\ep \to
0$. The coefficient $a_{4}(\xi)$ is, on the other hand, singular which violates
the status of IR perturbative finiteness of massless quenched QED$_3$. The
obtained singularity is such that all of its gauge-fixing dependent terms are
entirely determined by lower order contributions in agreement with the
properties of the LKF transformation.

In closing, let's note that the four-loop singularities were found to contribute
to the coefficient $a_{4}(\xi)$ with a factor $\sim (1-\xi)$ and, thus,
$a_{4}(\xi)$ is finite in the Feynman gauge. The reason for this intriguing
effect is not clear at present and its elucidation requires additional research.

\section*{Acknowledgments}
One of us (A.V.K.) thank A.\ G.~Grozin and A.\ L.~Kataev for useful discussions.
The work of V.P.G.\ is supported by the National Academy of Sciences of Ukraine
(project 0116U003191) and by its Program of Fundamental Research of the
Department of Physics and Astronomy (project No. 0117U000240). The work of
A.F.P.\ is supported by the Foundation for the Advancement of Theoretical
Physics and Mathematics ``BASIS.''

\printbibliography
\end{document}